\documentclass[12pt]{iopart}
 
\usepackage{verbatim}
\usepackage{soul}
\usepackage{xcolor}

\usepackage{graphicx, epstopdf} 

\newcommand{\curl}{\nabla \times}
\newcommand{\dver}{\nabla \cdot}

\newcommand{\der}[3][]{ \frac{ \partial^{#1}{#3} }{ \partial{#2}^{#1} } }
\newcommand{\fder}[3][]{ \frac{ d^{#1}{#3} }{ d{#2}^{#1} } }
\newcommand{\vect}[1]{\mathbf{#1}}

\begin{document}

\title{A critical examination of relativity theories}

\author{Max Tran}
\ead{mtran@kbcc.cuny.edu} 
\address{Department of Mathematics, Kingsborough Community College, Brooklyn, NY 11235-2398, USA, Phone: 718-368-6707}

\vspace{10pt}


\begin{abstract}
We give a brief critical examination of the special theory of relativity and a similar Newtonian framework to the first order of the $v/c$ ratio, focusing on the phenomena of aberration, Fresnel dragging, and the Doppler effect. We will show that both frameworks yield the same equations modeling these phenomena up to first order. We do this at a level understandable to anyone with a working knowledge of calculus so that the underlying ideas do not get lost in abstruse mathematical formulations. 
\end{abstract}



\section{Introduction}

Richard Feynman, the gifted teacher and Nobel laureate in Physics, said the following in a recorded lecture on the Scientific Method:
\begin{quotation}
\noindent If it disagrees with experiments, it's wrong. In that simple statement is the key to science. It doesn't make any difference how beautiful your guess is, it doesn't matter how smart you are who made the guess, or what his name is ... If it disagrees with experiments, it's wrong. That's all there is to it.
\end{quotation}
Replacing `guess' with `theory' would not impact the thrust of his statement. In this spirit, we will examine how well various relativity theories model reality, focusing on the three phenomena of aberration, Fresnel dragging, and the Doppler effect. In our view, a relativity theory is a system of equations that relates the equations modeling some process in one system of reference to another set of equations modeling the same or similar process in another system. We will examine these theories at a level understandable to anyone with knowledge of calculus so that the underlying ideas do not get lost in abstruse mathematical formulations of Minkowski space, tensors, or geometric algebra. We will examine these theories for logical consistency and agreement with physical principles. A similar approach can be found in the book of O'Rahilly. \cite{rahill}

To begin, we give some operational definitions of space and time to keep in mind when examining theories that rescale their coordinates.  After thousands of years of debates and investigations, we can say that space and time are subtle relations between events used to organize our perceptions of reality and are not definable in terms of other concepts. We may appeal to our experiences to get a sense of their meaning and mathematically model them with properties like stretchability, curvature, or flow, but that does not mean they actually have these properties. In practical terms of measurement, there are several operational meanings to both space and time. In one sense, space is the separation between two points, measured by counting how many standard length units it takes to span the separation. In another sense, space is a framework used to assign coordinates to events for our analysis. In such a framework, we may use various metrics, measures of separation, with different curvature properties to assign the distance between two points.  As to time, it has two practical definitions: duration and date. Duration is measured by counting the number of cycles of some periodic process: the swings of a pendulum or the number of cycles of an electronic oscillator controlled by observing the spectrum line of some atom. In essence, a clock is just a counter combined with an oscillator. As to date, it is just the duration from some arbitrary zero, and once this arbitrary zero is chosen, a duration can be found as the difference of two dates. In this sense, a duration is a time interval between the start and end of a process.  

It is an accepted principle of science that \textbf{under the same conditions}, physical processes should not depend upon dates or positions, but only on the duration of interactions and relative displacements. This principle is often obscured by the common practice of assigning the start of a process to be the zero date. Since choosing the zero of the date parameter to be the start of an interaction is not always possible when several processes are unfolding, we will not engage in this custom. And to avoid the confusion between duration and date, we will always use these two words instead of `time.' For the most part, we will use a framework of space with the Euclidean metric and a date parameter that is the same throughout all of space at any given moment. Some may take this as the absolute space and absolute time of Issac Newton. In a sense, it is, but this framework is just an idealization to make our analysis as simple as possible and is no more real than a framework of spacetime that curve, with clocks needing to be synchronized. Of course, in the analysis of real experiments, we need to be careful and take into account that clocks may have different rates of oscillations depending upon their motions, positions in a gravity well, or surrounding environments. We will also use an isotropic medium in all our models so that the speed of any wave is the same in all directions.

In the next section, we will look at the equations of the relativity theories of Poincare, Lorentz, and Einstein--namely the Lorentz transformation and transposition. In the third section, we will show a derivation of the Lorentz transposition to first order using the Newtonian framework of subrelativity. In the fourth section, we will derive the aberration formula from the transposition formula of subrelativity. The fifth section will show a derivation of the Fresnel dragging coefficient similar to what is done with the Lorentz transposition. In section six, we will derive the formula for the Doppler effect and examine its treatment by the special theory of relativity.  In section seven, we examine covariant and invariant versions of Maxwell's equations. In the final section, we conclude with some indications of future studies.

\section{Poincare, Lorentz, and Einstein's Special Relativity}

Around the turn of the twentieth century, many physicists worked on what would later be called relativity theory, including the likes of Woldemar Voigt, George FitzGerald, Joseph Larmor, Henri Poincare, and Hendrik Lorentz. In 1887, an obscured journal published Voight's article on a set of transformations similar to the Lorentz transformation that preserved the form of one version of the wave equation. Early in 1889, Oliver Heaviside used his version of Maxwell's equations to show that the electric field from a moving spherical charge distribution would appear contracted in the direction of motion. Some months later, FitzGerald published the conjecture that bodies in motion are contracted to explain the null result of the 1887 Michelson and Morley aether-wind experiment. In 1892, Lorentz independently presented the same idea in a more detailed manner. Lorentz and Larmor also looked for the transformation under which one version of Maxwell's equations retained their forms when transformed from the aether, a stationary reference frame, to a moving frame. They extended the Fitzgerald--Lorentz contraction hypothesis by modifying the date coordinate to what Lorentz called ``local time.'' Their works were generally known to researchers before 1905, the year that Albert Einstein published his paper on relativity. 

We will focus on Einstein's special relativity theory since it is easy to formulate and is widely accepted. This special theory of relativity is based on two hypotheses, in his own words:\cite{einstein1923}
(1) The laws by which the states of physical systems undergo change are not affected, whether these changes of state be referred to the one or the other of two systems of coordinates in uniform translatory motion.
(2) Any ray of light moves in the `stationary' system of coordinates with the velocity $c$, whether the ray be emitted by a stationary or by a moving body.

Hypothesis (1) is usually called the relativity principle and is implicit in the work of Newton, as evidenced by his words:\cite{newton}
\begin{quote}
The motion of bodies included in a given space are the same among themselves, whether that space is at rest or moves uniformly in a straight line without circular motion.
\end{quote}
This hypothesis allows us to define inertial frames and is often taken to mean that there is no preferred inertial frame. (In practice, one frame may be more convenient to work with.) Another way to interpret this principle is the impossibility of determining the uniform motion of an inertial system using experiments confined to that system. To detect its motion relative to outside references, detectors must be able to access these outside references. Einstein and many others take this principle to mean that the equations describing a phenomenon have the same form in all inertial frames, so-called covariancy, or form invariance. Poincare may have been the first to enunciate this interpretation of the relativity principle.\cite{whit} We will examine this interpretation in more detail below.  One should contrast covariance with invariance, the property of being unchanged under some transformation. In common physics usage, the adjective covariant may sometimes be used informally as a synonym for invariant. 
For instance, the Schrodinger equation does not keep its form under the transformations of special relativity. Thus one might say that the Schrodinger equation is not covariant. In contrast, the Dirac equation has the same form in any coordinate frame of special relativity, so one might say that it is covariant. But it is more precise to say that the Dirac equation is invariant and that the Schrodinger equation is not, but this is not the usual practice. Yet the Dirac equation is not invariant under the transformations of general relativity, nor is it in any sense covariant. Thus proper use should indicate the transformation under which invariance or covariance is considered.

Hypothesis (2) is called the invariancy of light speed. This constancy is often taken to hold for all inertial frames of reference or standards of rest. In this strong form, the hypothesis is unverifiable since we can not test it in all frames of reference. We can only take this as an axiom of our models and partially test the consequences of it, or reject it as we see fit. The weaker form of this hypothesis is that light speed is constant relative to the `medium' and so is independent of source motion. This hypothesis can be tested by experiments to some extent. And there have been experiments to test this weaker form of the second hypothesis, using elementary particles that emit light. These experiments compare the speed of light emitted when the particles are `at rest' compare to when they are moving and thus can only show that the speed of light is independent of source motion, not that light speed is the same in all inertial frames.\cite{lspeed,lspeed2} The Michelson-Morley and Kennedy-Thorndike types of experiments are often quoted to support the invariancy of light speed. These experiments often use reflected light to measure two-way light speed. They are not direct one-way measurements of light speed, so their results need careful analysis and are not free of controversy.

Later in his career, Einstein made some interesting statements concerning the domain of validity of his special relativity theory. For instance, in his 1913 paper ``Outline of the generalized theory'' \cite[V4, D13, p.153]{einsteincol} we find 
\begin{quote}
I have shown in previous papers that the equivalence hypotheses leads to the consequence that in a static gravitational field the velocity of light
$c$ depends on the gravitational potential. This led me to the view that the special theory of relativity provides only an approximation to reality; it should apply only in the limit case where differences in the gravitational potential in the space-time region under consideration are not too great.
\end{quote}
The implication is that light speed is not constant but is only approximately so for regions with a nearly uniform gravitational field. These statements seem to suggest that the medium for light transmission, its standard of rest, is the dominant gravitational field in the region of space under observation. Petr Beckmann proposed a similar theory in his book {\em Einstein Plus Two.}

In terms of fitness to experiments, the various relativity theories of Poincare, Lorentz, and Einstein are equivalent since they all deduce the Lorentz transformation and its consequences. Einstein once said: ``... the Lorentz transformations, the real basis of the special relativity theory.''\cite{einstein1935} Thus, any experiments that support the consequences of these equations are unable to distinguish between these theories or any theory which produces the same equations. Now, there are many ways to obtain the Lorentz transformation, but we will only give the light-sphere approach that uses basic algebra and the above two hypotheses. A similar approach to this appeared in Einstein's 1920 book {\em Relativity: the special and general theory}.\cite{einstein1920} For a concise derivation using Newton's laws and the covariancy of a version of Maxwell's equations, see Dunstan's article.\cite{dunstan}

Consider two system $S$ and $S'$ with $S'$ moving with constant velocity $v$ relative to $S$ along its $x$-axis. Let the two systems' origins $O$ and $O'$ coincide at the zero date, $t = t' = 0$.
Let a spherical signal with speed $c$ relative to an isotropic medium be emitted from the origin of $S$ at the date $t=0$ and is received at the date $t$ at the point  $(x,y,z)$ in $S$. The equation of the spherical volume enclosed by the wavefront in $S$ is then: 
\begin{equation} \label{stat}
	x^2 + y^2 + z^2 - c^2 t^2 = 0.
\end{equation}
Since Einstein and many people take the relativity hypothesis to mean that the equations in all inertial frames should have the same forms, the equation of the wavefront in $S'$ should have the form like the one above. According to them, the equation for the wavefront in $S'$ at $t'$ is
\begin{equation} \label{moving}
	 x'^2 +  y'^2 + z'^2 - c'^2 t'^2 = 0. 
\end{equation}
If the second hypothesis in either form is used, then $c' = c$. But we will show that this hypothesis is not needed to get the Lorentz transformation, so for now, we keep the distinction between $c$ and $c'$ and note that $c'$ is just an unknown parameter. 

Since there are no movement along the $y$ and $z$ axis, $ y  = y' $ and $ z  = z' $, by subtracting (\ref{stat}) and (\ref{moving}) and rearranging, we get:
\begin{equation} \label{combined}
	x^2 - c^2 t^2 =  x'^2  - c'^2 t'^2.
\end{equation}
Now suppose the position and date coordinates in the $S'$ frame is related to the coordinates in the $S$ frame by a linear transformation:
\begin{eqnarray}
	x' &= \alpha x + \beta t,  \label{lt1} \\
	t' &= \gamma x + \delta t.  \label{lt2}
\end{eqnarray} 
There are many reasons to assume the transformation is linear, but the best one may be to try the simplest form first.
At the instant $t$, $O'$ is at the position $x= v t$ with respect to $S$. Putting this equation and $x' = 0$ into (\ref{lt1}) gives
\begin{eqnarray}
	0 &= \alpha v t + \beta t, \nonumber \\ 
	\beta &= - \alpha v.  \label{lt3}
\end{eqnarray}
At the instant $t'$, $O$ is at the position $x' = -v t'$ with respect to $S'$. Putting this equation and $x=0$ into (\ref{lt1}) gives,
\begin{equation}
	-v t' =  \beta t. 
\end{equation}
Combining this with (\ref{lt2}) when $x=0$ yield
\begin{eqnarray}
	-v \delta t &=  \beta t, \\ 
	\quad \beta &= - \delta v.  \label{lt4}
\end{eqnarray}
Equations (\ref{lt3}) and (\ref{lt4}) implies that $\alpha = \delta$. Replacing $\delta$ by $\alpha$ and $\beta$ by $- \alpha v$ in (\ref{lt1}) and (\ref{lt2}), we get 
\begin{eqnarray} \label{ltpart}
	x' &= \alpha ( x - v t), \\
	t' &= \gamma x + \alpha t. \nonumber
\end{eqnarray}
Putting these values for $x'$ and $t'$ into (\ref{combined}), we obtain
\begin{equation} \label{lt5}
	x^2 - c^2 t^2 = \alpha^2 (x- vt)^2  - c'^2 (\gamma x + \alpha t)^2.
\end{equation}

Equation (\ref{lt5}) must be true for all $x$ and $t$ and since they are independent variables,  the coefficients of $x^2$, $xt$ and $t^2$ on both sides must be equal. This fact yield three equations for the three unknowns $\alpha$, $\gamma$ and $c'$:
\begin{equation}
	\alpha^2 - \gamma^2 c^2 = 1, \qquad 	\alpha^2 v + \alpha \gamma c'^2 = 0,   \qquad 	\alpha^2 v^2 - \alpha^2 c'^2 = -c^2. 
\end{equation} 
Solving these equations yield
\begin{equation} \label{results}
	\alpha = \frac{1}{\sqrt{1 - v^2/c^2}}, \qquad	\gamma = \frac{-v/c^2}{\sqrt{1 - v^2/c^2}},   \qquad	c' = c. 
\end{equation} 
Putting these results into (\ref{ltpart}) gives the Lorentz transformation
\begin{equation} \label{ltF1}
	x' =  \frac{x - v t}{\sqrt{1 - v^2/c^2}}, \quad 	y' = y, \qquad 	z' = z, \qquad	t' = \frac{t - vx/c^2}{\sqrt{1 - v^2/c^2}}. 
\end{equation}
Hermann Minkowski, the mathematician who first formulated special relativity using four-vectors, had this to say:
\begin{quotation}
\noindent For these equations invariance [covariance] under a Lorentz transformation is a purely mathematical fact, which I will call the Theorem of Relativity. This theorem essentially depends
upon the form of the differential equation for the propagation of waves with the velocity of light.
\end{quotation}
We just proved this theorem using the covariancy of the light sphere equations.

As others have done before, we note that the value of $c$ never entered into the derivation of the Lorentz transformation, abbreviated as LT. The implication is that it can be applied to any wave moving through some medium. Since Lorentz derived the LT from electromagnetism where $c$ is the speed of light, this value for $c$ is the one most often used. Also note that the LT rescaled the position and date coordinates, but since physical events do not depend upon arbitrary coordinates, what should be used to express the result of experiments is the Lorentz \textbf{{\em transposition}}, so-called by O'Rahilly, in either its finite-difference or differential form:
\small
\begin{equation} \label{ltF3}
\fl \qquad	\Delta x' =  \frac{\Delta x - v \Delta t}{\sqrt{1 - v^2/c^2}}, \qquad \Delta y' = \Delta y, \qquad \Delta z' = \Delta z, \qquad \Delta t' = \frac{\Delta t - v \Delta x/c^2}{\sqrt{1 - v^2/c^2}},  
\end{equation}
\normalsize
where the $\Delta$ quantities represents relative displacements or duration. Although the LT and the Lorentz transposition have the same structures, it should be noted that the LT expresses a particular relationship between arbitrary coordinates of position and date, while the Lorentz transposition expresses the same relationship between displacement and duration of an event. The LT implies the Lorentz transposition, but the reverse is not true in general. To see this, try to derive the Lorentz transposition using similar reasoning as above when the emitter is at $(x_0,0,0)$, and the starting date is $t_0$, then see if the LT follows without additional assumptions. 
The Lorentz transposition is used to derive the relativistic velocity addition formula, which yields $c$ as an upper limit on speed. Since $c$ is often taken to be light speed, many take this to mean that nothing can travel faster than light. But another way to interpret the relativistic speed limit is that the covariancy of the equations is no longer true at speeds greater or lesser than $c$.  The above derivation shows that the Lorentz transformation is an algebraic consequence of requiring a spherical wavefront in one frame to appear as a spherical wavefront in another frame and even force the speed of the wave to be the same in both frames to ensure covariancy. The contraposition is that a speed greater or lesser than $c$ in one frame implies that the equations are no longer covariant under the LT. There have been many published articles of experiments on signals or forces whose actions travel faster than light speed. Check the references \cite{faster1,faster}. 

A careful examination of the above light-sphere derivation of the Lorentz transformation brings some questions to mind. Does the relativity principle mean or imply covariancy under the LT? In particular, does a spherical wavefront in one inertial frame need to appear as a sphere in another inertial frame? Regarding the first question, many often take the relativity principle to mean that the phenomenon within an isolated system should depend only upon the relative position, relative velocity, and relative
acceleration of its parts. Others take it to mean covariancy of certain equations under the LT. According to Whittaker, it was Henri Poincare, a preeminent mathematician, and physicist, who put forth this interpretation, which unfortunately or fortunately became widely accepted. Poincare said `According to the Principle of Relativity the laws of physical phenomena must be the same for a ``fixed'' observer as for an observer who has a uniform motion of translation relative to him ...' \cite[v 2 p 30]{whit} 
I think it is a mistake to equate the covariancy of certain equations under the LT with the principle of relativity.  Even in electromagnetism, covariancy is only true by redefining the magnetic and electric fields in the moving system to preserve the forms of one version of Maxwell's equations.\cite{covarMax} 
Because the field transformation seemingly gives rise to a magnetic field in the detector rest frame from a purely electric field in a frame moving relative to the detector, some people believe that magnetic fields are due to relativity. As if identifying a mathematical model with physical reality will make our beliefs true. O'Rahilly outline this condition with the quip: ``Best described as mental autointoxication superinduced by an overdose of metaphors.'' \cite[p 739]{rahill} But Maxwell's equations and the wave equations have several forms, not all covariant under the LT, as we will demonstrate in a later section. This begs the question as to which version of Maxwell's equations should be covariant and what form the transformation must take for covariancy.

Regarding whether a spherical wavefront in one inertial frame appears as a sphere in another, let us apply the LT to a sound wave or an underwater light sphere that is spreading outward from a stationary central point at a speed of $u$. The natural choice for the origin is at the emitter, and the equation for the wavefront is a sphere $x^2 + y^2 + z^2 = u^2 t^2$.
Applying the inverse LT $x = \gamma (x' + vt')$, $t=\gamma (t' + v x'/c^2)$ for a frame that is moving with velocity $v$ along the $x$ axis, we get an equation that is supposed to describe the wavefront in a moving frame $\gamma^2 x'^2(1 - u^2v^2/c^4) + y'^2 + z'^2 + 2x' v t' \gamma^2 ( 1 - u^2/c^2) = \gamma^2 t^2 (u^2 - v^2)$, which is not a sphere unless $u=c$. Thus the equation describing the wavefront in these systems are not covariant under the LT.

There is a mistake in the argument from (\ref{moving}) to (\ref{lt5}) to get the Lorentz transformation that invalidates it. Richard Waldron remarked in his book that physically the light-sphere should be centered at the stationary emitter in $S$ and not at the origin of the moving frame.\cite[p 75--6]{waldron} Thus (\ref{moving}) is incorrect. According to $S'$ the emitter is at the position of the origin in $S$, at $x' = -v t'$  on the date $t'$, so the equation of the light sphere centered at the emitter in $S'$ that would replace (\ref{moving}) should really be 
\begin{equation} \label{moving2}
	 (x' + v t')^2 +  y'^2 + z'^2 - c'^2 t'^2 = 0.
\end{equation}
Reworking the derivation again with the above equation and (\ref{ltpart}) yield the equation
\begin{equation} \label{ltA5}
	x^2 - c^2 t^2 = [ \alpha (x- vt) + v (\gamma x + \alpha t)]^2  - c'^2 (\gamma x + \alpha t)^2,
\end{equation}
instead of (\ref{lt5}). This is true for all $x$ and $t$, so by expanding the above equation and collecting the coefficients of like terms we get the following equations:
\begin{equation} \label{result}
	\alpha^2 + 2 \alpha v \gamma  + v^2 \gamma^2 - \gamma^2 c'^2 = 1, \qquad -2 \alpha \gamma c'^2 = 0,  \qquad   \alpha^2 c'^2 = c^2. 
\end{equation}
Solving the above equations yield
\begin{equation} \label{results2}
	\alpha = 1, \qquad 	\gamma = 0,  \qquad	c' = c, 
\end{equation} 
giving us a very familiar transformation, often called the Galilean transformation:
\begin{equation} \label{gt}
	x' = x - v t, \qquad  y' = y, \qquad z' = z, \qquad t' = t.
\end{equation}
Thus the light-sphere covariancy derivation, if done correctly, cannot give us the Lorentz transformation. Similar derivations based upon Einstein's synchronization procedure and light-cone likely have the same defect. This does not however invalidate other derivations of the LT. This derivation does however show that the light sphere equation is covariant under the Galilean transformation, which gives the usual velocity addition law. Thus from the receiver perspective, different parts of the wavefront are moving away from their positions at different speeds, with some being greater than light speed.

Many researchers claim a large body of experimental data supports the Lorentz transposition and its consequences. We will examine these experiments to see if this is the case. In the next section, we will derive a formula similar to the Lorentz transposition to the first order of $v/c$ from a Newtonian framework called subrelativity so that all experiments to first-order would also support this approach. We would need to examine experiments accurate to the second-order of $v/c$ to distinguish between special relativity or the Newtonian framework of subrelativity.

\section{Corelativity, Interrelativity, Subrelativity}

This section is a concise summary of O'Rahilly's deep and insightful analysis of relativity found scattered throughout chapter 9 of his book.\cite{rahill} He came up with three divisions of relativity, which he called corelativity, interrelativity and subrelativity. O'Rahilly also derived the transformations and transpositions between such systems concerning the emission and reception of a wave through an isotropic medium. He made the observation that experiments rest on the possibility of isolating the system on which we are experimenting. Ideally, any outside influence on such a system is either zero or irrelevant to the phenomenon under study. In his words:
\begin{quote}
A complete system is thus that totality of physical objects whose behavior is wholly determined by internal factors and is independent of what is happening elsewhere. At first sight it would seem obvious that the phenomena inside such a system are independent of such an extrinsic relationship as the rate at which the system as a whole is changing its distance with respect to outside objects. The point cannot be decided by a priori kinematical reasoning, however; it is experience which shows us that a set of objects which has an acceleration relative to the fixed stars is not strictly an isolated system ... {\em An isolated physical system must therefore be such that its acceleration (relative to a Newtonian framework) is either zero or negligible. It is then found that all phenomena internal to such a system are independent of its motion, with constant velocity in a straight line, relative to any other system.} We call this the principle of \textbf{{\em corelativity}}. \cite[p 428]{rahill}
\end{quote}
A completely isolated system must include its own medium--air, water, or elastic solid--for any wave phenomenon dependent upon them so that the principle of corelativity applies not only to projectiles but also to waves. Let $E_1, R_1, M$ be the emitter, receiver, and medium in the system $S_1$ that is nonaccelerating, and let $E_2, R_2, M$ be the emitter, receiver, and medium in the system $S_2$ that is moving in relation to $S_1$ with constant velocity $v$. Whether the propagation is ballistic or medium-based, measurements in each system are identical if done under the same conditions. 
\begin{figure}
\begin{center}
{\includegraphics[scale=1.2]{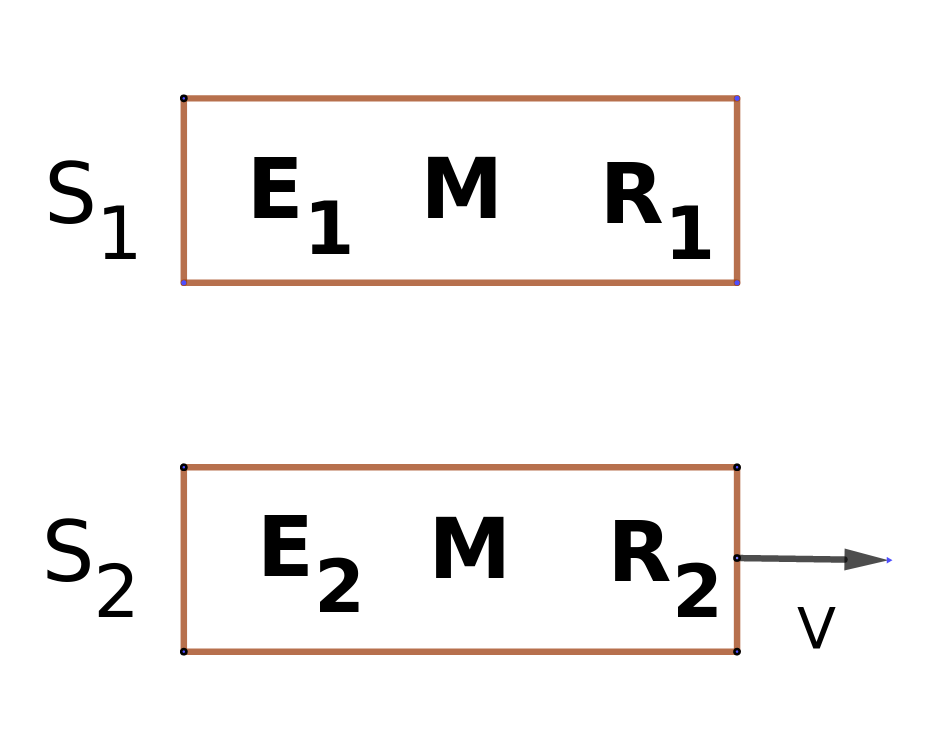}}
\caption{Corelativity: The systems $S_1$ and $S_2$ are completely enclosed containing their own emitters, receivers, and any medium required to carry the signal. $S_2$ is moving with velocity v relative to $S_1$. The transformation relating events between the two systems is the identity.} \label{fig:corel}
\end{center}
\end{figure} 

\begin{figure}
\begin{center}
{\includegraphics[scale=1.2]{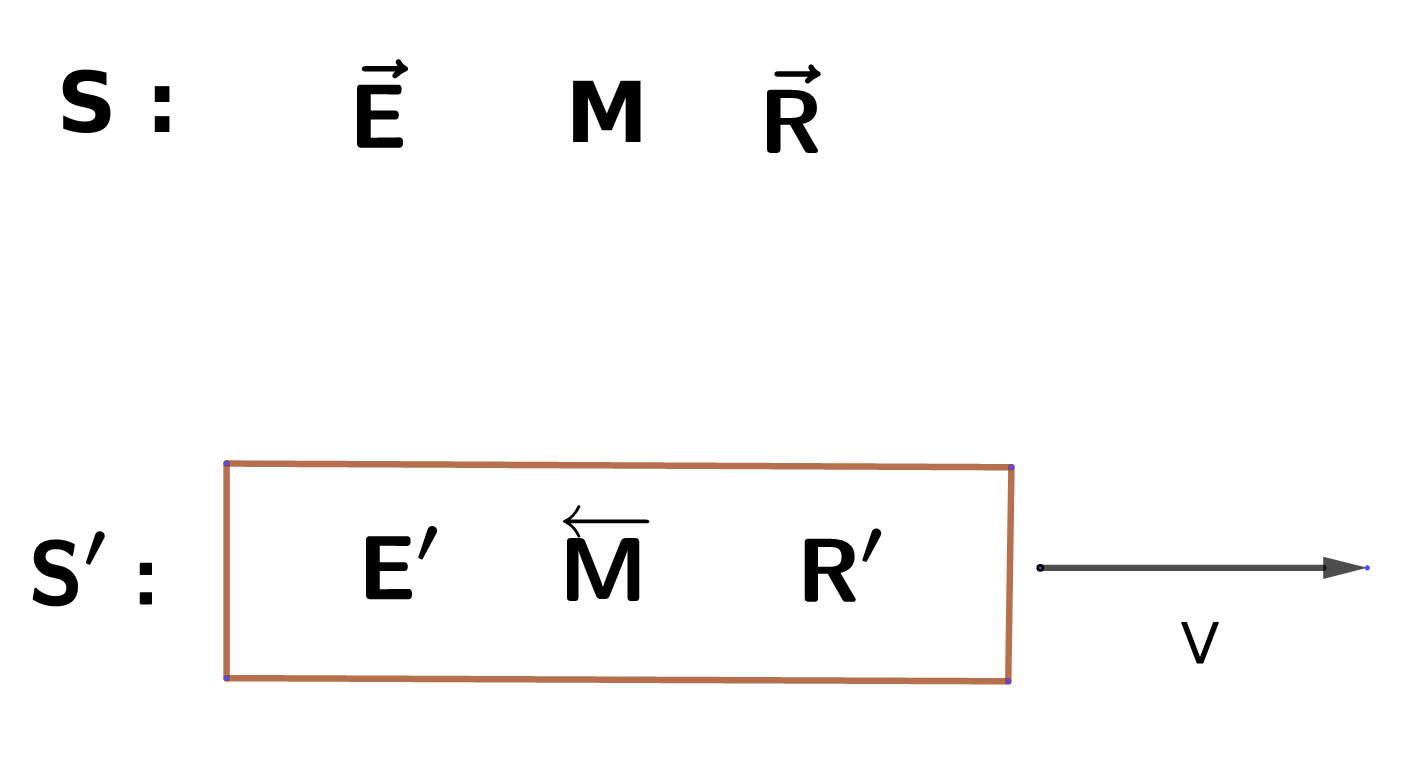}}
\caption{Interrelativity: In system $S$, the emitter and receiver are moving with velocity $v$ through the medium. In system $S'$, $E'$ and $R'$ are at rest in the system, and the enclosed medium is moving with velocity $-v$ relative to it, with the whole system moving at velocity $v$ with respect to $S$.} \label{fig:interrel}
\end{center}
\end{figure}

Now consider a system $S$ having an emitter and receiver moving with velocity $v$ through an isotropic medium $M$ and an isolated system $S'$ with an emitter and receiver at rest in $S'$, and a medium in $S'$ moving with velocity $-v$ relative to it. The system $S'$ as a whole is moving with uniform velocity $v$ relative to $S$. These two systems are said to be \textbf{{\em interrelative}}. Now by the principle of corelativity, phenomena internal to $S'$ are independent of the movement of $S'$ as long as it is moving at a constant velocity. The relation between the above two systems' space and date coordinates is called \textbf{ {\em interrelativity} }.  Figure \ref{fig:interrel} illustrate the situation.

\begin{figure}
\begin{center}
{\includegraphics[scale=1.2]{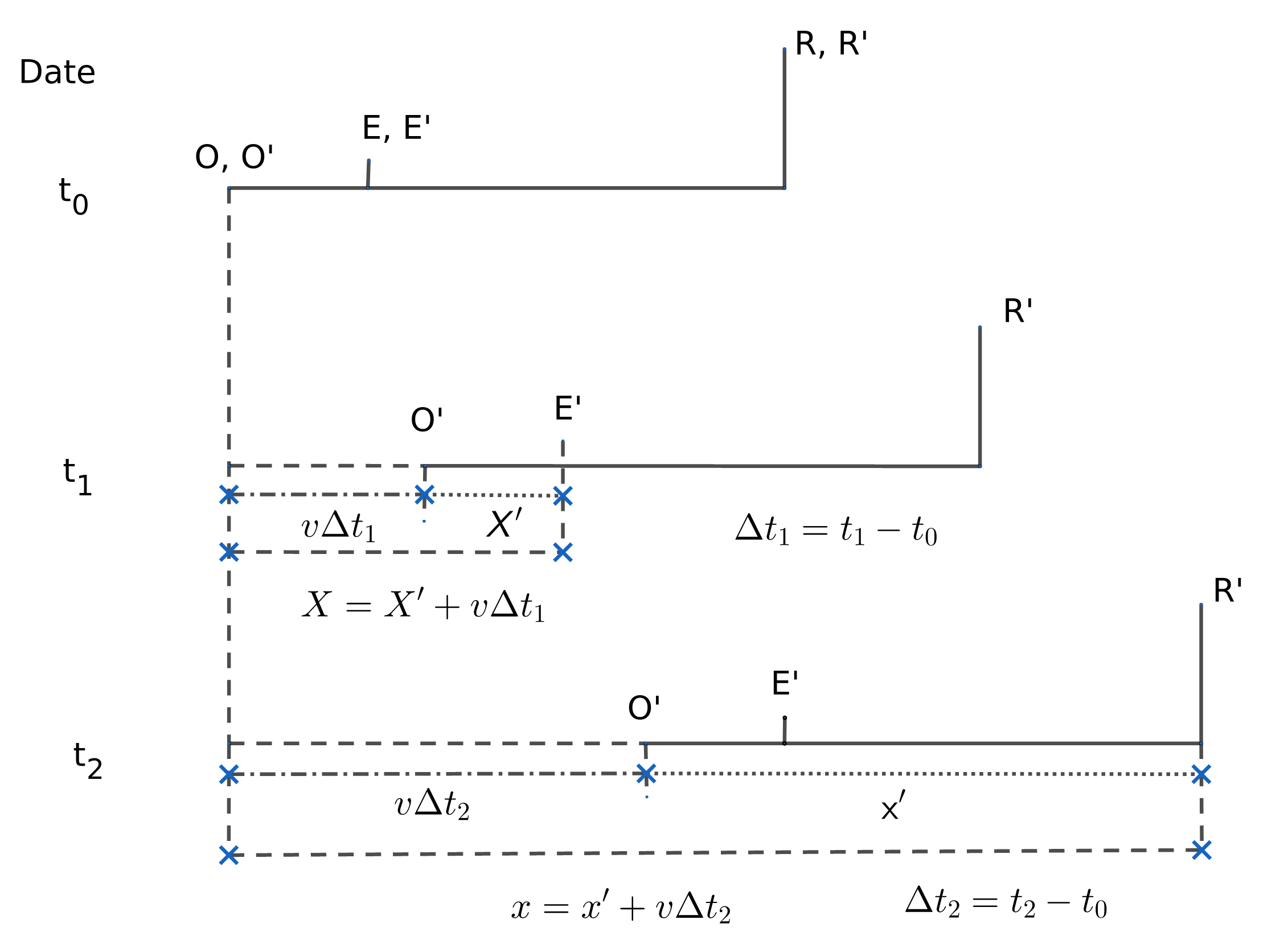}}
\caption{Transposition of Interrelativity: In system $S$ the emitter and receiver are moving with velocity $v$ parallel to the $x$-axis through the medium. In system $S'$, $E'$ and $R'$ are at rest in the system, with the whole 
system moving through the medium at velocity $v$. The transformation between the two systems is often called the Galilean transformation.} \label{fig:interrelSys}
\end{center}
\end{figure}

The transposition of interrelativity for the duration and displacement has the form $\Delta x' = \Delta x - v \Delta t$, $\Delta y' = \Delta y$, $\Delta z' = \Delta z$ and $\Delta t' = \Delta t$, which many called the Galilean transformation, and can be easily obtained from figure \ref{fig:interrelSys}. Note that this transposition is independent of any value of the wave-velocity $c$, much less the value $c= \infty$.

Next we consider two systems: (1) system $S_o$ in which the emitter and receiver are at rest in the medium $M$ and (2) system $S_m$ in which the emitter and receiver move with constant velocity $v$ through the same medium, with $c$ being the wave speed relative to the medium. These two systems are said to be \textbf{{\em subrelative}}. It should be noted that the two systems are not identical but do involve the same mode of transmission in the same medium. Also, note that $v$ is not some arbitrary velocity but is a quantity that is internally relevant to the system $S_m$ and that $S_m$ is not self-contained.
\begin{figure}
\begin{center}
{\includegraphics[scale=1.2]{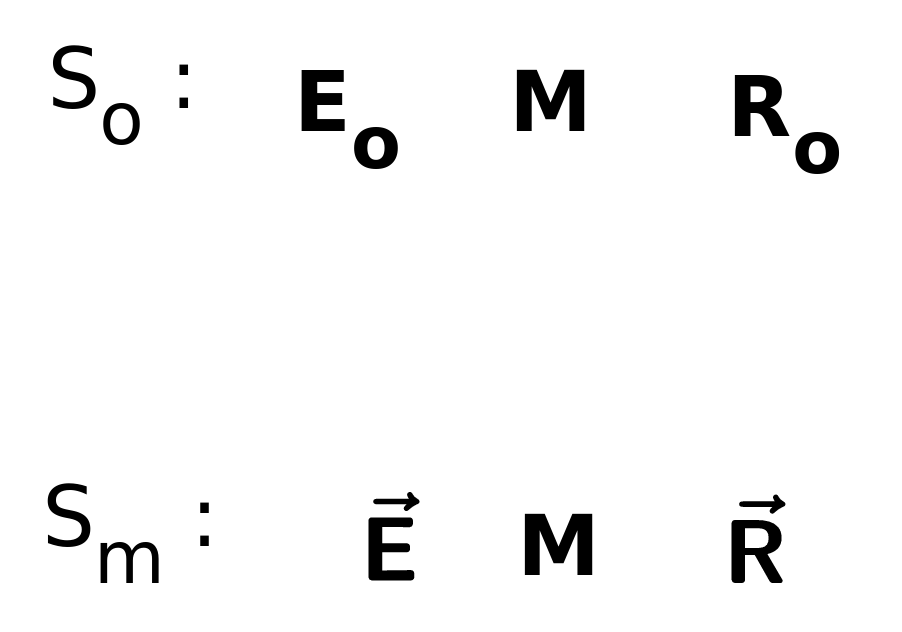}}
\caption{Subrelativity: In system $S_o$, $E_o$ and $R_o$ are at rest in 
the medium $M$, while in $S_m$, the emitter and receiver are moving with velocity $v$ through the same medium.} \label{fig:subrel}
\end{center}
\end{figure} 

\begin{figure}
\begin{center}
{\includegraphics[scale=3]{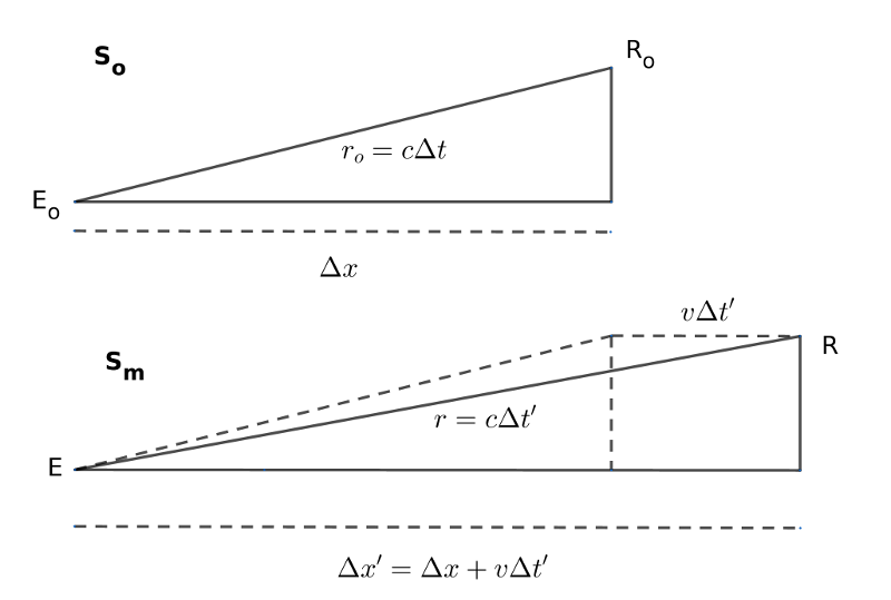}}
\caption{Subrelative systems: In $S_m$ the emitter and receiver are moving with velocity $v$ through the medium. In system $S_o$, $E_o$ and $R_o$ are at rest in the system, with the whole system at rest in the medium.} \label{fig:subrelSys}
\end{center}
\end{figure}

We will now derived the transposition of subrelativity. At the emission date $t_o$ we suppose that the two systems coincide, with $E_o$ and $E$ both sending out a wave at the position $(x_o,y_o,z_o)$.  The receivers $R_o$ and $R$ are at $(x,y,z)$. At the reception date $t$ in $S_o$, we get the transmission duration $\Delta t = t - t_o$ and the equation for the distance covered by the wave $(x - x_o)^2 + (y - y_o)^2 + (z - z_o)^2 = c^2 \Delta t ^2$. See figure \ref{fig:subrelSys}. At the reception date $t'$ in $S_m$, the transmission duration is $\Delta t' = t' - t_0$. When the wave reached $R$ in $S_m$, this receiver has advanced $v \Delta t'$ along the x-axis to the position $(x', y', z')$. It should be clear that $x' = x + v \Delta t', y' = y, z' = z$, and that the distance covered by the wave in $S_m$ is $(x' - x_o)^2 + (y' - y_o)^2 + (z' - z_o)^2 = c^2 \Delta t'^2$. Thus we get the equations:
\begin{eqnarray*} 
	( x' - x_o - v \Delta t')^ 2 + (y' - y_o)^2 + ( z' - z_o)^2 &= c^2 \Delta t^2 , \\ 
					(x' - x_o)^2 + (y' - y_o)^2 + (z' - z_o)^2  &= c^2 \Delta t'^2 .
\end{eqnarray*}
Setting $\Delta x' = x' - x_o$, the relation between the two transmission-durations is then 
\begin{eqnarray}
	\Delta t^2 &= ( 1 + v^2/c^2) \Delta t'^2 - 2 v \Delta x' \Delta t'/c^2 , \nonumber \\ 
	\qquad  &=  (\Delta t' - v \Delta x'/c^2)^2 + (v/c)^2 (\Delta t'^2 - \Delta x'^2/c^2).  
\end{eqnarray} 
For $v$ much smaller than $c$, $v^2/c^2$ will be close to zero and so to first-order of $v/c$, the duration becomes
\begin{equation}
	\Delta t =  \Delta t' - v \Delta x'/c^2,
\end{equation} 
which some may recognized as the `local time' of Lorentz, but it is just the transmission duration of a pulse in $S_o$.  The complete set of equations relating the displacement and duration of events in the two subrelative systems is
\begin{equation} \label{E:subrel}
\fl \quad	\Delta x = \Delta x' - v \Delta t', \quad \Delta y = \Delta y', \quad \Delta z = \Delta z', \quad \Delta t = \Delta t' - v \Delta x'/c^2.
\end{equation}
The inverse subrelative transposition to the same order is 
\begin{equation} \label{E:Invsubrel}
\fl \quad	\Delta x' = \Delta x + v \Delta t, \quad \Delta y' = \Delta y, \quad \Delta z' = \Delta z, \quad \Delta t' = \Delta t + v \Delta x/c^2.
\end{equation}
To first-order of $v/c$, the gamma factor in the Lorentz transposition is approximately 1, $(1 - v^2/c^2)^{-1/2} \approx 1$. Thus the subrelative transposition (\ref{E:subrel}) is identical to the Lorentz transposition to the first-order, except that the speed $v$ can not be an arbitrary number, but must be the speed of the receiver with respect to the medium. In other words, the Lorentz transposition and inverse transposition rescale the position and time coordinates by a gamma factor, which just represent the length contraction hypothesis, but is otherwise just the transposition of subrelativity.

\begin{figure}
\begin{center}
{\includegraphics[scale=1.7]{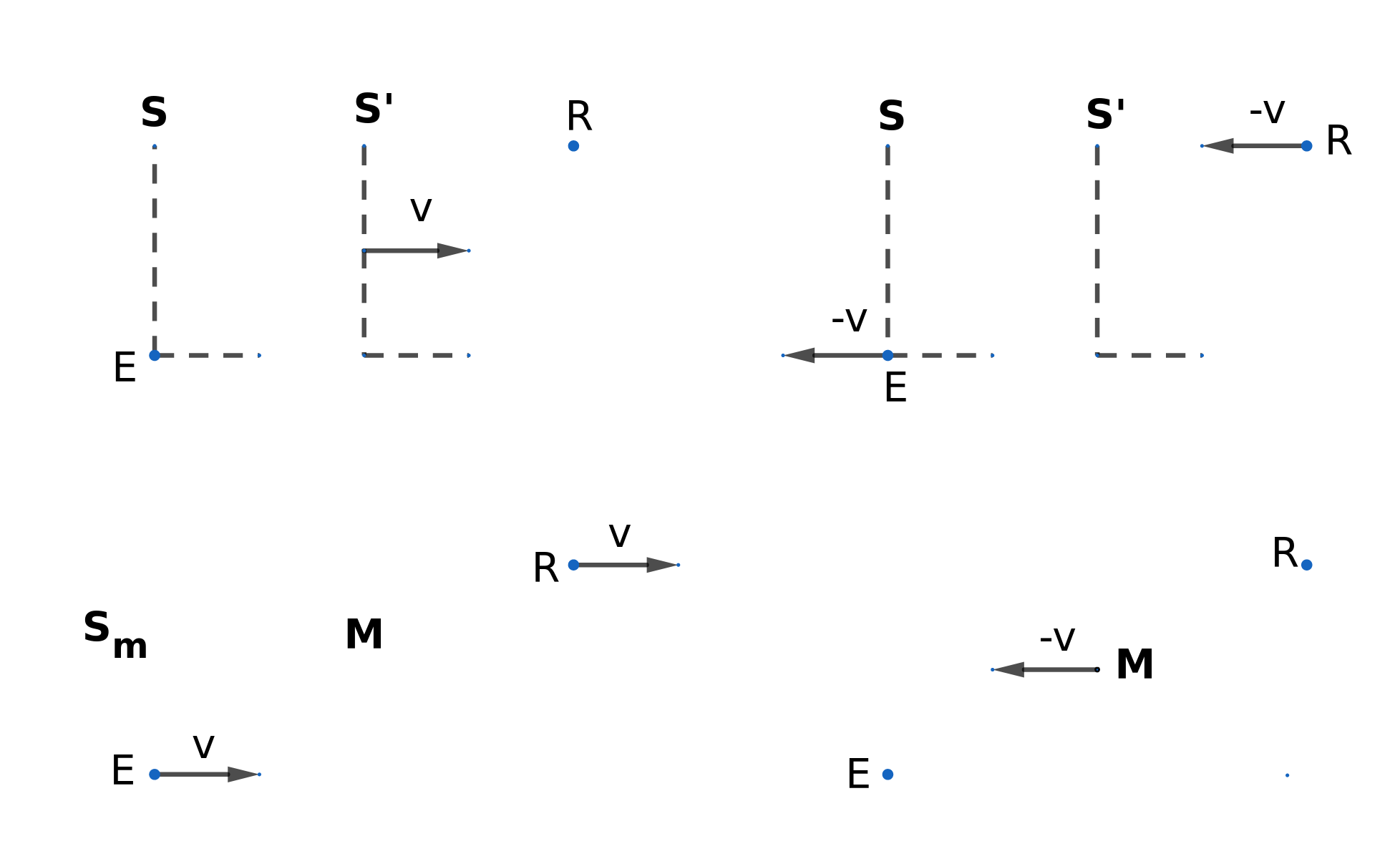}}
\caption{The systems in the light-sphere and subrelative derivations can be viewed from two perspectives. The top frames used in the Special Theory of Relativity: (1) $S'$ has velocity $v$ relative to the emitter and receiver $E$ and $R$; or (2) $S'$ is stationary with $E$ and $R$ moving with velocity $-v$ relative to $S'$. 
In the $S_m$ system of subrelativity: (1) $E$ and $R$ are moving with velocity $v$ through the medium; or (2) the medium is moving with velocity  $-v$ relative to $E$ and $R$.} \label{fig:SubComp}
\end{center}
\end{figure}

Attentive readers may note the sign difference between the transpositions of Lorentz and subrelativity. In the light-sphere derivation, the LT relates two systems $S$ and $S'$ moving relative to each other with speed $v$. The standard picture is that $E$ and $R$ are stationary in $S$ and $S'$ is moving to the right with speed $v$ relative to them. With respect to $S'$ considered as stationary, $E$, $R$, and $S$ have velocity $-v$ relative to it. In the subrelative $S_m$ system, $E$ and $R$ both have velocities $v$ relative to the medium $M$. If we take the system $S_m$ along with $E$ and $R$ as the standard of rest, then the medium has velocity $-v$ relative to them. The medium plays the same role as $S'$ in the above light-sphere derivation but with $-v$ velocity. Thus $-v$ in the subrelative derivation corresponds to $v$ in the light-sphere derivation. The two transposition formulas would match exactly had we set up the subrelative system with $E$ and $R$ moving to the left along the $x$-axis relative to $M$. See figure \ref{fig:SubComp}. 

It should be clear that we derived the above formulas for subrelativity using Newtonian physics, with no particular reference to electromagnetism or optics, so they are just as applicable to sound as to particles scattering. The only role that the medium played in our model is that of the standard of rest for the velocities. Also, note that there is a physical reason for the speed limit of the subrelative model. If the speed of the receiver, $v$, relative to the medium exceeds the medium speed of the signal, $c$, the receiver will never get any wave from the emitter! Thus the model would not be applicable for $v \geq c$. This does not mean that the receiver must have a speed less than $c$ or that $c$ is the ultimate speed limit. 

Lorentz, in 1895, may have been the first to apply the first-order formula of subrelativity in electromagnetic theory to show that Maxwell's equations are covariant to the first-order.\cite{lorentz}
According to Darrigol, a historian of science, Lorentz and Cohn were able to derive the equations that model aberration, Fresnel dragging, and the Doppler effect using the first-order Lorentz transposition.\cite{darri1995} We will show below that the transposition of subrelativity would yield the same equations. In fact, any experiment used to support the Lorentz transposition to the first-order of $v/c$ would also support the transposition of subrelativity since they are essentially the same equations.

\section{Aberration}

Aberration, also called velocity aberration, is a phenomenon wherein objects appear displaced towards the direction of motion of the observer compared to when the observer is stationary. It is historically significant because of its role in the development of the theories of light and the special theory of relativity. Astronomers began to notice this effect in the late 1600s, and in 1727 James Bradley gave a classical explanation in terms of the finite speed of light relative to the orbital motion of the Earth around the Sun.

\begin{figure}
\begin{center}
{\includegraphics[scale=1.7]{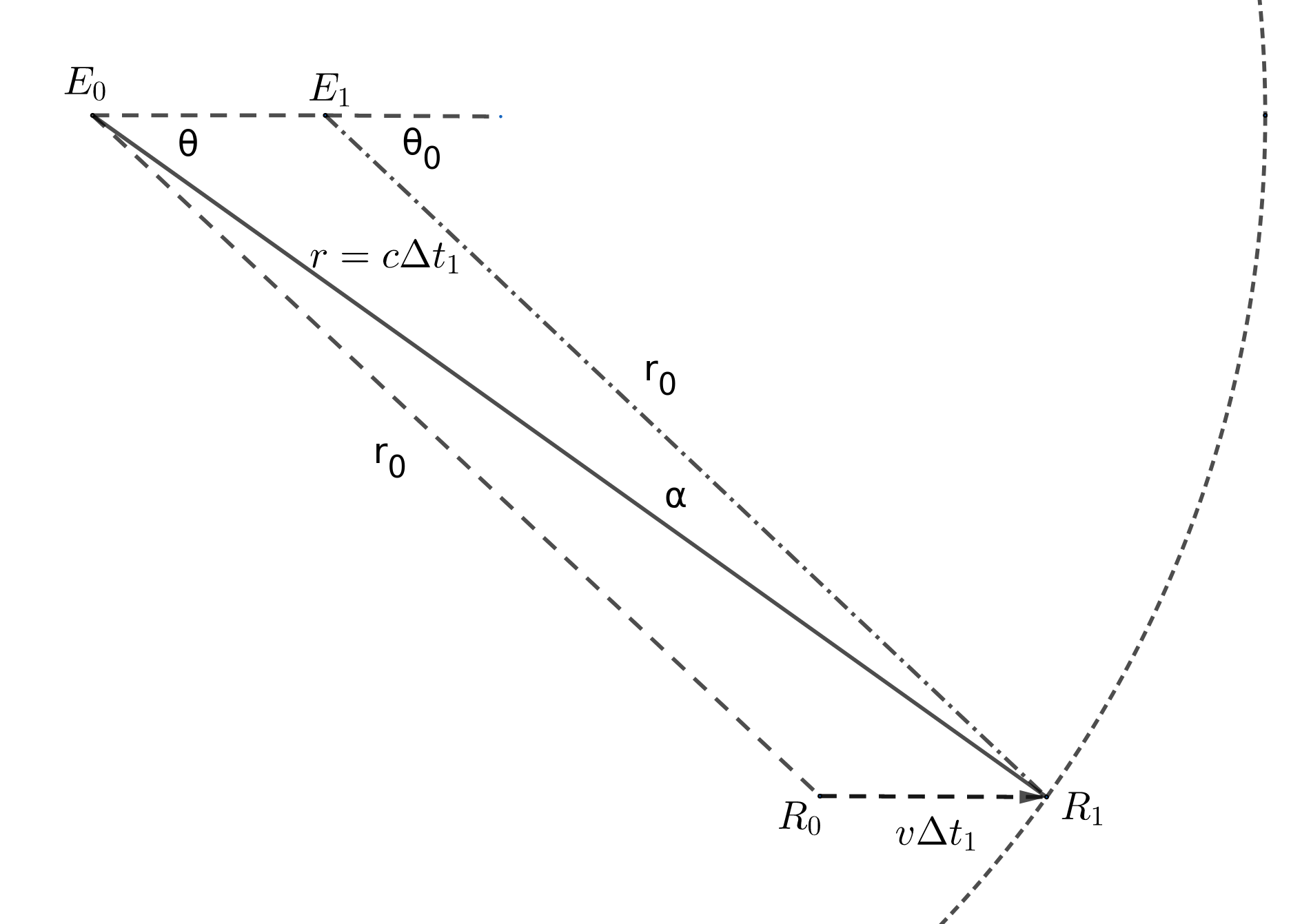}}
\caption{Modeling aberration with subrelative systems. In one system, the emitter and receiver, $E_o$ and $R_o$, are at rest in the medium. In the other system, $E_1$ and $R_1$ are moving parallel to each other with velocity $v$ through the medium. In the depicted scenario, a signal is emitted at $E_o$ when the two systems coincide and then received at $R_1$ when the second system has moved.} \label{fig:aberation}
\end{center}
\end{figure}

Let us examine aberration using the subrelativity framework. To do this we first generalize the equations of subrelativity to an arbitrary position in three-dimensional space. Referring to figure \ref{fig:aberation}, let the displacement vector $E_oR_o$  have direction cosines $(l_o,m_o,n_o)$ and magnitude $r_o = c \Delta t$ and let $E_oR_1$ have direction cosines $(l,m,n)$ and magnitude $r = c \Delta t_1$. The receiver, $R$, in system $S_m$ is still moving parallel to the $x$-axis. To first-order, we have the equations
\begin{eqnarray}
	t &= t_o ( 1 + l_o v/c), \nonumber \\
	r &= r_o ( 1 + l_o v/c), \nonumber \\
	l &= \frac{\Delta x}{r} = \frac{ \Delta x_o + v r_o/c}{ r_o (1 + v l_o /c) }  = \frac{ l_o + v/c}{ 1 + l_o v /c }, \\
	m &= \frac{\Delta y}{r}  = \frac{ m_o}{ 1 + l_o v /c}, \\
    n &= \frac{\Delta z}{r}  =  \frac{ n_o}{ 1 + l_o v /c}. 
\end{eqnarray}
Inversely, to first-order we have
\begin{equation}
 l_o =  \frac{ l - v/c}{ 1 - l v /c }, \qquad	m_o = \frac{ m}{ 1 - l v /c}, \qquad n_0 =  \frac{ n}{ 1 - l v /c}.
\end{equation} 
For the two-dimensional case, figure  \ref{fig:aberation}, we have
\begin{eqnarray}
	\cos \theta = \frac{ \cos \theta_o + v/c}{ 1 + v \cos \theta_o /c}, \qquad	\sin \theta = \frac{ \sin \theta_o }{ 1 + v \cos \theta_o /c},
\end{eqnarray}
where $\theta$ is the angle of a light ray when the receiver is moving, and $\theta_o$ is the angle of the ray when the receiver is stationary. Thus the aberration angle is 
\begin{eqnarray}
\fl \quad	\alpha = \theta_o - \theta  &\approx \sin(\theta_o - \theta) = \frac{v \sin \theta_o /c }{ 1 + v \cos \theta_o/c} = \frac{v}{c} \sin \theta_o - \frac{v^2}{c^2} \sin \theta_0 \cos \theta_o + - \cdots, \nonumber \\
 & \approx \frac{v}{c} \sin \theta_o. \label{aberform}
\end{eqnarray}
Since $\theta_o$ and $\theta$ are dependent upon the date of observation, the aberration angle is also dependent upon the date.
Referring to figure \ref{fig:aberation}, with the signal emitted at $E_o$, $E_oR_o$ is the absolute (medium) path when $R_o$ is stationary relative to the medium, while $E_oR_1$ is the absolute path when the receiver is moving. $E_1R_1$ would be the absolute path if the receiver was at rest in the medium at $R_1$ and the signal emitted at $E_1$. Since $E_1R_1$ and $E_oR_o$ are parallels, $E_1R_1$ is the relative (apparent) path for the signal emitted at $E_o$ according to the moving receiver. Thus the emitter at $E_o$ seems to be at $E_1$, by tracing the signal relative path back to its apparent source. In summary, we arrived at the aberration formula using a subrelative system by identifying the absolute path in system $S_o$ with the relative path in $S_m$.

Note that the velocity $v$ that determines the aberration angle in the above model is the velocity of the receiver/observer relative to some medium and is independent of the source velocity. In truth, the medium in the above model only serves as the standard of rest, a frame of reference for the velocities. In the case of stellar aberration as observed on Earth, what is the medium or reference frame for the velocity? We will return to this question after looking at the special relativity theory treatment of this phenomenon. 

The usual relativistic derivation use the Lorentz transposition and some algebra to get the formula:
\begin{eqnarray}
	\alpha = \frac{v \sin \theta_o /c }{ \sqrt{1 - v^2/c^2} }.
\end{eqnarray}
A concise derivation using special relativity can be found in \cite{sr4p}. Taking the first-order approximation in $v/c$, yield the same formula as (\ref{aberform}), except that the velocity $v$ has a different interpretation, often depending upon who derived the formula.

As documented by Naur, there is a state of confusion among textbooks, journal articles, and scientists about the standard of rest for the velocity $v$, in essence, the meaning of the velocity.\cite{naur}  Naur traced this confusion back to the work of Einstein and classified the various derivations into three categories:
\begin{enumerate}
\item The velocity $v$ is the velocity of the observer relative to the source of light, found in the works of Einstein and many relativistic derivations.
\item The velocity $v$ is the velocity of one observer relative to another observer at a different date or direction of motion, called the ``two-observers theory'' and can either be based upon a Newtonian or a relativistic framework.
\item The classical derivation is based upon a preferred frame, which in the past was identified with the aether and is essentially the approach used by Bradley.
\end{enumerate}

In the first approach, the velocity $v$ would depend upon the rotational speed of the Earth and the speed of the star relative to some common frame. This would lead to varying aberration angles for rotating binary stars, as pointed out by Eisner.\cite{binary} Since this is not observed, this interpretation of $v$ is simply wrong. 
In the second approach, one observer serves as the standard of rest for the other observer, so in essence, define a reference for the velocity $v$ and $c$. The second observer can be the original observer at a state of rest.
In the third approach, the preferred frame is often taken to be the medium required for the transmission of light waves after the particle theory of light was abandoned. The medium is just the standard of rest for the velocity $v$ and $c$, so there is no practical difference between the second and third approaches if they agree upon a reference frame for the velocities. To calculate the biggest contribution to stellar aberrations on Earth, $v$ is the Earth's orbital velocity relative to a frame in which the Sun or the barycenter of the solar system is at rest. A much smaller contribution due to the Earth's rotation is accounted for when the rotational velocity is taken into account.

In summary, all of these approaches use different justification to obtain the same first-order formula (\ref{aberform}), albeit with different interpretations of $v$. Some of these justifications are ad-hoc and unverifiable or even simply wrong. Yet experiments can only determine if the aberration formula (\ref{aberform}) describes reality and not the correctness of the reasoning behind the various derivations. To summarize, these various derivations give us a formula to use and say that aberration is caused by the motion of the observer relative to something. Nevertheless, we think the explanation is incomplete without some statements as to why the distant stars appeared to move due to the motion of the observer/receiver. We think the apparent motion is due to tracing the relative path of light back to the apparent position of the source, akin to tracing the relative path back to an image in the phenomenon of refraction.  

\section{Fresnel Dragging}

We will now give a derivation for the Fresnel coefficient using the framework of subrelativity similar to the approach that uses the Lorentz transposition. For the system $S_o$, let $V_o = c/n$ be the speed of a wave with respect to the medium of refractive index $n$ and $\Delta t_o$ be the transmission duration.
Let $r_o = V_o \Delta t_o$ and $(l_o, m_o, n_o)$ be the direction cosines of $\vect{r}_o$ in $S_o$.
Let $V$, $\Delta t$,  $r = V \Delta t$ and $(l, m, n)$ be the corresponding quantities in the moving system $S_m$ of a subrelativity framework moving parallel to the $x$-axis of $S_o$, with velocity $v$. We want a formula that relates $V$ to $V_o$. Previously, we derive the subrelative transposition when both $r^2 - c^2 \Delta t^2 = r_o^2 - c^2 \Delta t_o^2$ were zero, we now assume that it also hold when neither are zero: 
\begin{equation}
 \Delta x = \Delta x_o + v \Delta t_o,  \qquad \Delta t = \Delta t_o + v \Delta x_o/c^2.
\end{equation}
By definition and the above equations, we have
\begin{eqnarray*}
	l = \frac{\Delta x}{r}  = \frac{ \Delta x_o + v \Delta t_o}{V ( \Delta t_o + v \Delta x_o /c^2)}  = \frac{V_o}{V} \frac{ l_o + v/V_o}{ 1 + l_o v V_o /c^2 }.
\end{eqnarray*}
Similarly we get
\begin{equation}
	m = \frac{\Delta y}{r}  = \frac{V_o}{V} \frac{ m_o}{ 1 + l_o v V_o /c^2}, \qquad n = \frac{\Delta z}{r}  = \frac{V_o}{V} \frac{n_o}{ 1 + l_o v V_o /c^2}.
\end{equation}
By squaring and adding these equations, then keeping only terms of the first order in $v/c$ we get
\begin{equation}
	 \frac{V^2}{V_0^2} =  \left( 1 + \frac{ 2 l_o v}{V_o} \right) \left( 1 - \frac{2 l_o v V_o}{ c^2} \right) = 1 + 2\frac{l_o v}{V_o} \left( 1 - \frac{V_o}{c^2} \right).
\end{equation}
Take squaring root of both side and substituting $V_o = c/n$, we get the Fresnel coefficient formula:
\begin{equation}
	 V =   \frac{c}{n} + l_o v \left( 1 - \frac{1}{ n^2} \right).
\end{equation}

We have just shown that the Fresnel coefficient formula is an algebraic consequence of assuming that the first-order subrelative transposition formula with medium transmission speed $c$ applies to a medium of transmission speed $c/n$. Anyone familiar with relativistic literature may see that using the Lorentz transposition for deriving the Fresnel coefficient does not prove the coefficient $1 - 1/n^2$. The process just assumes it--in the assumption that the Lorentz transposition is applicable.

\section{The Doppler Effect}

The Doppler effect is the shift in wave frequency caused by the relative motion between a wave source and the wave receiver/observer. The drop in pitch of a passing siren is an example of this effect that many have experienced. Christian Johann Doppler first described the effect in 1842 as the process wherein the frequency of starlight increase or decrease due to the relative motion of the star. It has many important applications in astronomy and various technologies, including radars, medical imaging, flow management, velocity profile measurement, and satellite communication.  
We will model this phenomenon using a Newtonian framework, then compare this model to the special theory of relativity treatment of this effect.

\begin{figure}
\begin{center}
{\includegraphics[scale=1.2]{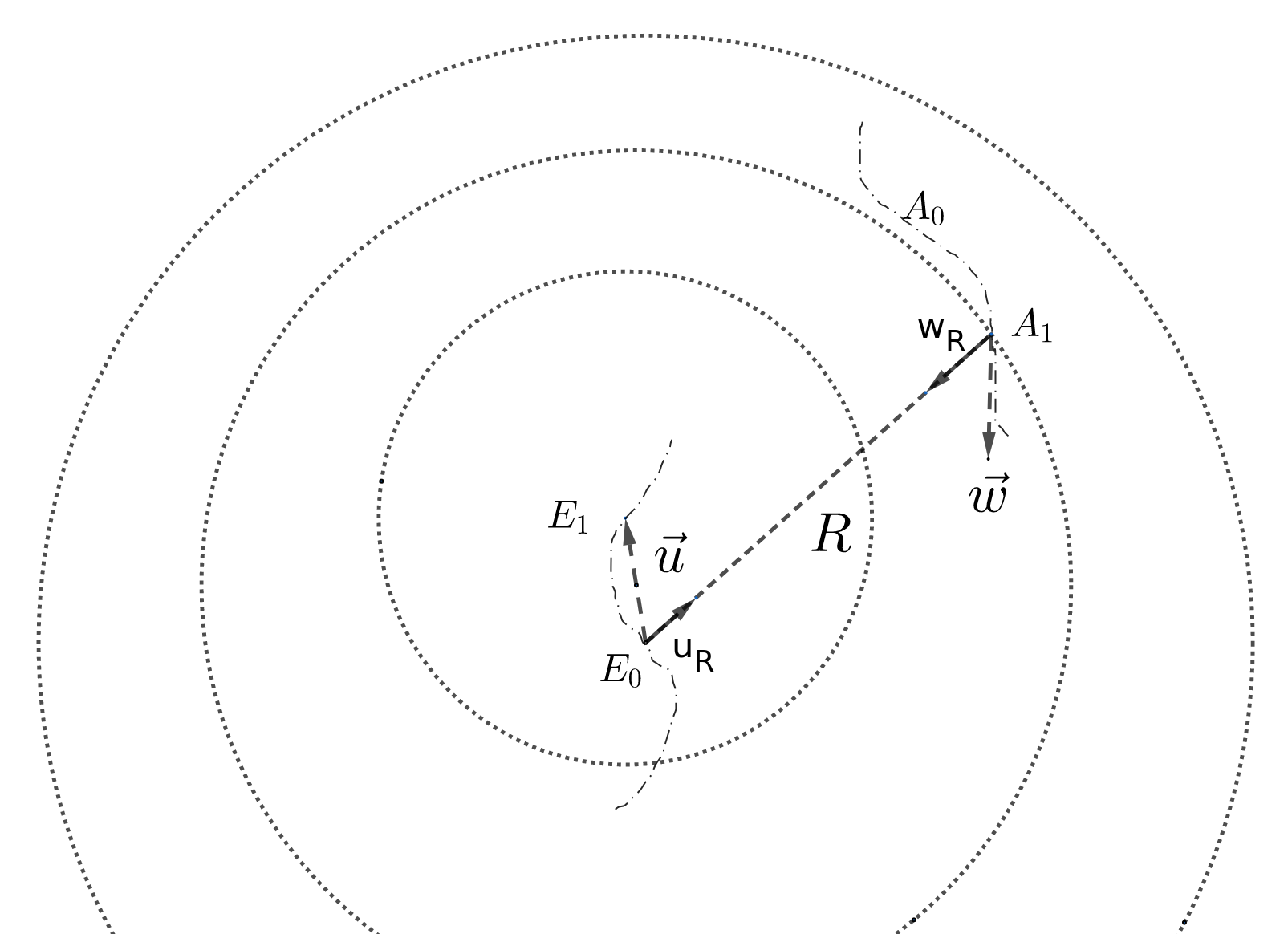}}
\caption{Scenario for the doppler effect: $E_o$ and $E_1$ represent the positions of the emitter at the date of emission and reception of the first wave front, while $R_0$ and $R_1$ represent the positions of the receiver at those same dates. The segment $E_oR_1$ is the join between the emitter and receiver along which their relative radial velocity $w_r - u_r$ is computed.} \label{fig:doppler}
\end{center}
\end{figure}

Let an emitter, $E$, and receiver, $R$, be moving with constant velocities $\vect{u}$ and $\vect{w}$ with respect to an isotropic medium. Let $ X = X_o + u_x T, ... x = x_o + w_xt, ...$ be the coordinates of $E$ at emission date $T$ and of $R$ at reception date $t$ respectively. 
Let the segment $ER$ joining the positions of the emission and reception events have length $r$ and direction-cosines $(l,m,n)$, then
\begin{equation}
	r^2 = \sum_i [ x_{oi} + w_it - (X_{oi} + u_i T) ]^2,
\end{equation}
so that
\begin{equation} \label{dplr1}
	\frac{dr}{dT} = \sum_i l_i \left( w_i \frac{dt}{dT} - u_i \right) = w_r \frac{dt}{dT} - u_r,
\end{equation}
where $u_r$ and $w_r$ are the radial velocities of the emitter and receiver along their join.
Since $r= c (t - T)$,
\begin{equation} \label{dplr2}
	\frac{dr}{dT} = c \left( \frac{dt}{dT} - 1 \right).
\end{equation}
By equating the right hand side of (\ref{dplr1}) and (\ref{dplr2}) then solving for $dt/dT$, we get  
\begin{equation}
	 \frac{dt}{dT} = \frac{ 1 - u_r/c}{ 1 - w_r/c}.
\end{equation}
See figure \ref{fig:doppler}.
If $f_E$ is the frequency of emission and $f_R$ that of reception, then $f_E dT = f_R dt$, since the number of waves emitted in the interval $dT$ is equal to that received in the interval $dt$.
Hence
\begin{eqnarray}
\frac{f_R}{f_E} &= \frac{1 - w_r/c}{1- u_r/c} = 1 - \frac{w_r - u_r}{c} \left( 1 - \frac{u_r}{c} \right)^{-1} \nonumber \\
 &= 1 - \frac{w_r - u_r}{c} - \frac{w_r - u_r}{c} \left( \frac{u_r}{c} \right) - \frac{w_r - u_r}{c} \left( \frac{u_r}{c} \right)^2 + \cdots
\end{eqnarray}
This formula is applicable to any form of wave motion in a medium and if we accept Maxwell's theory of electromagnetic waves, then it must apply to EM waves as well. From the above formula, we see that the biggest contributor to the Doppler effect is the relative radial velocity of the emitter and receiver, $w_r - u_r$, even when there is a medium carrying the signal. If the relative radial velocity is zero during the time interval between each wave, then according to this model there is no doppler effect. Even though the velocities $\vect{u}$ and $\vect{w}$ are constants, the angles they make with the join may vary and so would $w_r - u_r$.  Of course, if $\vect{w}$ and $\vect{v}$ vary with the date, the current model would require modification to take that into account. Lastly, we obtained the classical Doppler formula by analyzing the phenomenon in one frame of reference, the frame in which the medium is stationary, and did not need to use two or more frames moving relative to each other. Thus relativity theories are irrelevant to obtaining the classical Doppler effect formula.

Let us now examine the formula of the Doppler effect as derived from special relativity. Einstein derived a formula of this effect in his 1905 paper and 1907 paper using the Lorentz transposition and a major assumption. In his words,
\begin{quote}
\noindent As I showed in an earlier paper [referring to the 1905 paper], it follows from these principles that a uniformly moving clock, judged from the stationary system, goes slower than when judged by a co-moving observer. If $p$ denotes the number of beats of the clock in unit time for the stationary observer, $p'$ the corresponding number for the moving observer, then $p/p' = (1 - v^2/c^2)^{-1/2}$, or approximately ... The radiation from the ions of the canal-rays is to be regarded as a quickly moving clock, hence the above formula is applicable. But it must be observed that the frequency $p'$ (for the co-moving observer) is unknown, so that the formula is not directly amenable to experimental investigation. It is assumed, however, that $p'$ is also equal to the frequency [$p_s$] which the same ion in the stationary state emits or absorbs. \cite{einstein1907}
\end{quote}
In terms of our above notation, $p = f_R$, $p' = f'_R$ are the corresponding frequencies of the receivers in two separate systems, and $p_s = f_{SE} = f_{SR}$, the frequency at the emitter and receiver when both are stationary with respect to the medium.
His derivation and most of the derivation that follows afterward have the form: $f_R = f'_R (1 - v^2/c^2)^{-1/2}$, by an application of the Lorentz transposition, then assume $f'_R = f_{SE}$, therefore $f_R = f_{SE} (1 - v^2/c^2)^{-1/2}$.  If the emitter and receiver also move, then the classical Doppler effect has to be taken into account and would modify the formula. 

The important difference is the gamma factor `correction' to the classical Doppler formula, often called the transverse Doppler effect or `time dilation' since it is present even when the relative radial velocity between the emitter and receiver is zero. To detect this gamma factor would require experiments to the second-order of $v/c$. The Ives and Stillwell type experiments are of this order, but they are not direct measurements since they use reflection. Direct tests use particle accelerators or the Mossbauer effect.  We will examine all of them when we compare theories at the second-order of $v/c$. In any case, it is the duration between each pulse of a wave train that is lengthened or shortened, not time itself. Needless to say, the existence of the transverse Doppler effect does not by itself prove the various postulates of SRT. Experiments would just confirm the formula, which can be obtained by assuming the Lorentz contraction and combining it with the transposition of subrelativity.

Any examinations of the doppler effect would be incomplete without mentioning its extension by Wladimir Michelson that lies outside the domain of any relativity theory. To the first-order, the Doppler shift formula is $f_R = f_E ( 1 - \dot{x}/c)$, where $x$ is the relative radial separation between the emitter and receiver. This relation holds also for rays undergoing reflections and refractions, as long as $x$ is replaced by the optical path $y= \sum_i n_i x_i$. The general formula for the Doppler-Michelson effect would then be
\begin{equation}
	f_R = f_E \left[ 1 - \sum_i \frac{( x_i \dot{n}_i + n_i \dot{x}_i)}{c} \right].
\end{equation}
The first term $\sum_i x_i \dot{n}_i $ account for the situation where media of uniform refractive indices move in the ray-path. Thus a change in frequency may be produced by a change in the thickness, density, or refractive indices of the media. M. A. Perot, in 1923, verified this formula in an experiment using twelve prisms rotated by an electric motor.\cite{perot} O'Rahilly gave a detailed analysis of the experiment in his book. It would be of value to repeat this experiment with more sensitive technology available today to get independent confirmation or refutation of this result.

\section{Covariant versus Invariant Maxwell's and Wave equations}

The Maxwell's equations in vacuum with a partial time derivative are the ones usually shown to be covariant under the Lorentz transformation, and have the form:
\begin{equation} \label{maxwell}
\fl \quad	\dver \vect{E} = - \frac{\rho}{\epsilon_o}, \quad \dver \vect{B} = 0, \quad \curl \vect{E} = - \der{t}{\vect{B}}, \quad \curl \vect{B} = \mu_0 \vect{J} + \frac{1}{c^2} \der{t}{\vect{E}}.
\end{equation}

Many people have worked on modifying or extending Maxwell's equations in various ways because the equations could not model all electromagentic or optical phenomena. Notable names among these are Heinrich Hertz, Oliver Heaviside and Emil Cohn. More recent revival of these modified equations can be found in the work of Phipps and others.\cite{chuby,phipps,pinheiro} These approaches often replace the partial time derivative in Maxwell's equation with the total time derivative or the convective derivative. (See the appendix for a derivation of the Maxwell-Faraday equation with the total time derivative). Phipps in particular showed that the Hertzian equations for EM are invariant under the Galilean (interrelative) transposition. The Hertzian EM equations in vacuum are as follow:
\begin{eqnarray} 
	\dver \vect{E} &= - \frac{\rho}{\epsilon_o}, \qquad \dver \vect{B} = 0, \nonumber \\
	\curl \vect{E} &= - \fder{t}{\vect{B}}, \qquad \curl \vect{B} = \mu_0 \vect{J}_m +  \frac{1}{c^2} \fder{t}{\vect{E}},  \label{hertzian}
\end{eqnarray}
where $\frac{d}{dt} = \der{t}{} + \vect{w} \cdot \nabla$ and $\vect{w}$ is the velocity of a test particle or as some take it the velocity of a moving reference frame.

The proof of invariancy under the interrelative transposition for the Hertzian EM equations readily follows from the Galilean velocity addition law and the following relation among derivatives between the two frames:
\begin{equation} 
	\nabla' = \nabla, \qquad \der{t'}{} = \der{t}{} + \vect{u} \cdot \nabla, 
\end{equation}
where $\vect{u}$ is the relative velocity between two interrelative systems. This relation can be proven by applying the chain rule on the interrelative transposition formula.

According to Darrigol, Hertz's version of Maxwell's equations ``brought Maxwell's theory to the highest degree of formal perfection.''\cite[p 257]{darri2}
If we accept that the Hertzian EM equations can fit more physical phenomena, shouldn't we require their covariancy under the Lorentz transformation? But they are not covariant, as we will show. Consider the standard pair of systems with one moving relative to another along their common $x$-axis with uniform speed $v$.
Using the inverse Lorentz transposition and the chain rule gives the following relations among the derivative operators: 
\begin{eqnarray} \label{ldiff}
	\der{x'}{} &= \gamma \left( \der{x}{} + \frac{v}{c^2} \der{t}{} \right), \quad \der{y'}{} = \der{y}{}, \nonumber \\
	\der{z'}{} & = \der{z}{}, \quad \der{t'}{} = \gamma \left( \der{t}{} + v \der{x}{} \right).
\end{eqnarray}
These can be used to show that (\ref{maxwell}) are covariant as long as the following transformations on the components of the fields are used:
\begin{eqnarray} \label{ltransf}
	E'_x = E_x, \quad E'_y = \gamma(E_y - v B_z), \quad E'_z = \gamma( E_z + v B_y), \\
	B'_x = B_x, \quad B'_y = \gamma \left (B_y + \frac{v}{c^2} E_z \right), \quad B'_z = \gamma \left( B_z - \frac{v}{c^2} E_y \right). \nonumber
\end{eqnarray}

Let $\vect{w}$ be the velocity of a test particle in the $S_o$ system. To show the noncovariancy of (\ref{hertzian}) under the LT, we just need to show that one equation does not retain it form after an application of (\ref{ldiff}) and (\ref{ltransf}). We will show that the equation
\begin{equation} \label{faraPrime}
	\nabla' \times \vect{E'} = - \fder{t'}{\vect{B'}} = -\der{t'}{\vect{B'}} - (\vect{w'} \cdot \nabla') \vect{B'},
\end{equation}
can not be transformed into
\begin{equation} \label{faraday}
	\curl \vect{E} = - \fder{t}{\vect{B}} = -\der{t}{\vect{B}} - (\vect{w} \cdot \nabla) \vect{B}.
\end{equation}
In fact, we just need to show that $(\vect{w'} \cdot \nabla') \vect{B'}$ does not transform into $(\vect{w} \cdot \nabla) \vect{B}$, since the remaining parts of the equations are just the covariant Maxwell's equations. 
By definition 
\begin{equation}
	\vect{w'} \cdot \nabla ' = w'_{x} \der{x'}{} + w'_{y} \der{y'}{} + w'_{z} \der{z'}{}.
\end{equation}
The relativistic velocity addition formula relate $\vect{w'}$ to $\vect{w}$ by:
\begin{equation} \label{speed}
w'_{x} = \frac{w_x - v}{ 1 - w_xv/c^2}, \quad w'_y = \frac{w_y}{\gamma ( 1 - w_xv/c^2)}, \quad w'_z = \frac{w_z}{\gamma (1 - w_xv/c^2)}.
\end{equation}
By combining (\ref{speed}) and (\ref{ldiff}), we get
\begin{eqnarray*}
\fl \quad \vect{w'} \cdot \nabla ' = \gamma \frac{w_x - v}{ 1 - w_xv/c^2} \left( \der{x}{} + \frac{v}{c^2} \der{t}{} \right) +\frac{w_y}{\gamma ( 1 - w_xv/c^2)} \der{y}{} + \frac{w_z}{\gamma (1 - w_xv/c^2)} \der{z}{},
\end{eqnarray*}
which will never equal $\vect{w} \cdot \nabla  = w_{x} \der{x}{} + w_{y} \der{y}{} + w_{z} \der{z}{}$ unless $v$ is zero. Thus the Hertzian EM equations are not covariant with respect to the Lorentz transformation.

Since the wave equation in some form is of great importance, let us briefly examine their covariancy. The scalar potential, vector potential, and the $\vect{E}, \vect{B}, \vect{D}, \vect{H}$ fields as modeled by the covariant version of Maxwell's equations all satisfy the covariant wave equation
\begin{equation} \label{coWave}
	\nabla^2 \psi  - \frac{1}{c^2}\der{t}\psi = f.
\end{equation} 
The corresponding wave equation in Hertzian's electrodynamics is
\begin{equation} \label{inWave}
	\nabla^2 \psi  - \frac{1}{c^2}\fder{t}\psi = f,
\end{equation} 
which can be easily shown to be invariant under the Galilean transposition. Whether it is covariant or even invariant under the Lorentz transposition is left as an exercise to the reader. 

Finally, it should be noted that the above wave equations are approximations in which the dampening terms involving a first-order time derivative are set to zero. Real waves are attenuated as they move thru a medium and lose their energy, however slowly. Thus models with no dampening term in the wave equation have a finite domain of validity, beyond which they lose their conformity to reality. In terms of electromagnetic waves, there must be some dampening to resolve Olbers' paradox. When the dampening terms are not set to zero, then the wave equation with either a partial or full time derivative will not be covariant under the Lorentz transformation, as was pointed out by Monti in his astute critical analysis of relativity.\cite{monti} The equations of physics exist to codify the connection between phenomena and takes whatever form they take according to the approximation used or simplifying assumptions made to model a given situation. To require that these equations retain their forms under some mathematical transformations and even equating such a requirement to be the relativity principle defies reason, in our opinion.

\section{Closing remarks}

Our examination of the special theory of relativity revealed its essence to be the Lorentz transformation (really transposition) and its algebraic consequences. The various relativity theories of Poincare, Lorentz, and Einstein obtained the Lorentz transformation by requiring the covariancy of certain equations in electromagnetism, namely some version of the wave equation or Maxwell's equations. We found that covariancy under the Lorentz transformation is a mathematical fact due to the structure of the equations and may or may not be true depending upon the equations.  We also found that requiring covariancy of the light sphere equation just places a speed limit beyond which or below which covariancy of the equation is no longer true. Thus the Lorentz transformation has nothing to do with the speed of light or the ultimate physical speed limit.  Perhaps the most important revelation is the existence of the subrelativity Newtonian framework that gives covariancy to the first-order of $v/c$. So that experiments to the first-order can not distinguish between the Lorentz transposition or the transposition of subrelativity. We would need to go to the second-order to see the differences between the two.

Future papers in this series will examine experiments, especially the Michelson moving mirror experiment, Michelson-Morley, Kennedy-Thorndike, Ives-Stillwell, and the Pi meson and Muon lifetime experiments, to see their fitness to the various relativity frameworks. 

\section{Appendix}

The proof of covariancy of one version of Maxwell's equations use the Maxwell-Faraday equation of the form:
\begin{equation} 
	\curl \vect{E} = - \der{t}{\vect{B}},
\end{equation}
which models induction from a time-varying magnetic field, but does not model induction due to motion.
Since the Maxwell-Faraday equation is usually derived from the law of electrical induction, let us derive a version that model both form of inductions. In terms of the induced electric field $\vect{E}$ and the inducing magnetic field $\vect{B}$, the Faraday induction law can be expressed by the equation:
\begin{equation} \label{intFara}
\oint_{\Sigma(t)}\vect{E} \cdot d\vect{s} = - \frac{d}{dt} \int_{S(t)} \vect{B} \cdot d\vect{A}, 
\end{equation}
where $\Sigma(t)$ is any closed path in space and $S(t)$ is any open surface bounded by $\Sigma(t)$. The path and surface may be moving in an arbitrary way or be stationary with the standard of rest being the frame in which $\vect{E}$ and $\vect{B}$ are measured. Applying Stokes' theorem to the left hand side of (\ref{intFara}), which requires certain smoothness assumption on the $\vect{E}$ field, gives the equation:
\begin{equation} \label{stokes}
	\oint_{\Sigma(t)} \vect{E} \cdot d\vect{s} = \int_{S(t)} \curl \vect{E} \cdot d\vect{A}.
\end{equation} 
Using a theorem from vector calculus on the total time rate of change of an integral over a moving surface with a velocity field $\vect{v} = \vect{v}(\vect{x},t)$ in the same coordinate system as $\vect{E}$ and $\vect{B}$, \cite[p 188]{kemmer} yield the equation:
\begin{equation} \label{derOfIntegr}
	- \frac{d}{dt} \int_{S(t)} \vect{B} \cdot d\vect{A} = - \int_{S(t)} \left[ \der{t}{\vect{B}} - \curl (\vect{v} \times \vect{B}) + ( \nabla \cdot \vect{B}) \vect{v}  \right] \cdot d\vect{A}.
\end{equation}
Equating the two right hand sides of (\ref{stokes}) and (\ref{derOfIntegr}), and bringing all the terms to one side under one integral gives:
\begin{equation} 
	\int_{S(t)} \left[ \curl \vect{E} + \der{t}{\vect{B}} - \curl (\vect{v} \times \vect{B}) + ( \nabla \cdot \vect{B}) \vect{v}  \right] \cdot d\vect{A} = 0.
\end{equation} 
This equation is true for any arbitrary surface $S(t)$ bounded by $\Sigma(t)$, thus implying the integrand equals zero and yielding the equation: 
\begin{equation}  \label{Fara}
	\curl \vect{E} + \der{t}{\vect{B}} - \curl (\vect{v} \times \vect{B}) + ( \nabla \cdot \vect{B}) \vect{v} = 0. 
\end{equation}
Using $\dver \vect{B} = 0$, and the identity 
\begin{equation}
\curl ( \vect{v} \times \vect{B}) = (\dver \vect{B} + \vect{B} \cdot \nabla) \vect{v} - (\dver \vect{v} + \vect{v} \cdot \nabla) \vect{B},
\end{equation}  
yield another form of the Maxwell-Faraday equation: 
\begin{equation}  \label{Max-Fara}
	\curl \vect{E} = - \der{t}{\vect{B}} - (\vect{v} \cdot \nabla) \vect{B} + (\vect{B} \cdot \nabla) \vect{v} - (\dver  \vect{v}) \vect{B}
\end{equation}
where $\vect{v}$ is the velocity of a test particle or point on the surface where the fields are measured. If $\vect{v}$ is independent of the space coordinates then the last two terms are zero and we end up with a secondary form of the Maxwell-Faraday equation:
\begin{equation} \label{Max-Fara2}
\curl \vect{E} = - \left[ \der{t}{\vect{B}} + (\vect{v} \cdot \nabla) \vect{B} \right] .
\end{equation}
Some researchers would identify the quantities inside the brackets as the total time derivative of $\vect{B}$ other as the convective derivative of $\vect{B}$.

Similar reasoning gives the Maxwell-Ampere law of the form:
\begin{equation} \label{Max-Amp}
\curl \vect{B} = \frac{1}{c^2} \fder{t}{\vect{E}} + \mu_o \vect{J},
\end{equation}
where $\vect{J}$ is the current from the perspective of a moving test particle. A legitimate concern to raise is the possibility of losing some important feature of our model in the conversion of an equation from an integral to a differential form. \\

\textbf{Acknowledgment} We would like to thank Steffen Kuhn whose questions and comments improved the paper.
\pagebreak



\begin{thebibliography}{99}

\bibitem{lspeed}
	T. Alvager, A. Nilson and J. Kjellman, ARKIV FOR FYSIK 26, 209--21 (1963)
	
\bibitem{chuby}
	A. E. Chubykalo and R. Smirnov-Rueda, Modern Physics Letters A 12, 1--24 (1997); DOI: https://doi.org/10.1142/S0217732397000029

\bibitem{darri1995}
	O. Darrigol, Am. J. Phys. 63, 908--15 (1995); DOI: http://doi.org/10.1119/1.18032
	
\bibitem{darri2}
	O. Darrigol, Electrodynamics From Ampere To Einstein (Oxford University Press, Oxford, UK 2000)	
	
\bibitem{dunstan}
	D. J. Dunstan, Phil. Trans. R. Soc. 366, 1861--5 (2008); DOI: https://doi.org/10.1098/rsta.2007.2195
	
\bibitem{einstein1907}
	A. Einstein, Annalen der Physik 23, 197--8 (1907)

\bibitem{einstein1920}
	A. Einstein, Relativity: The Special and the General Theory, Appendix I, Simple Derivation of the Lorentz Transformation. (1920)
	http://www.Bartleby.com/173/a1.html
		
\bibitem{einstein1923}
	A. Einstein, H. A. Lorentz, H. Weyl, and H. Minkowski, The principle of relativity : A collection of original memoirs on the special and general theory of relativity. (Dover, New York, NY 1952)

\bibitem{einstein1935}
	A. Einstein, Bull. Am. Math. Soc. 41, 223 (1935)
	
\bibitem{einsteincol}
	The collected papers of Albert Einstein. Princeton University Press, open-access website. 
	https://einsteinpapers.press.princeton.edu/

\bibitem{binary}
	E. Eisner, Am. J. Phys. 35, 817--9 (1967); DOI: http://doi.org/10.1119/1.1974259

\bibitem{kemmer}
	N. Kemmer, Vector Analysis: A physicist's Guide to the Mathematics of Fields in Three Dimensions (Cambridge Univeristy Press, Cambridge, UK, 1977)

\bibitem{faster1}
	S. Kuhn, Journal of Electromagnetic Analysis and Applications 12, 71-87 (2020); DOI: htts://doi.org/10.4236/jemaa.2020.126007
	
\bibitem{lorentz}
	H. A. Lorentz, Versuch einer Theorie der elektrischen und optischen Erscheinungen in bewegten Koerpern. (Brill, Leiden, NL, 1895)

\bibitem{monti}
	R. A. Monti, Physics Essays 9, 238 (1996); DOI: http://doi.org/10.4006/1.3029231

\bibitem{naur}
	P. Naur, Physics Essays 12, 358--67 (1999); DOI: https://doi.org/10.4006/1.3025390

\bibitem{newton}
	I. Newton, Principia leges motus cor 5 (Glasgow, UK 1871)
	
\bibitem{rahill}
	A. O'Rahilly, Electromagnetic Theory: A Critical Examination of Fundamentals (Dover, New York, NY, 1965)
	http://archive.org/details/ElectrodynamicsORahilly

\bibitem{perot}
	M. A. Perot, Comptes Rendus 178, 380--3 (1924)

\bibitem{phipps}
	T. E. Phipps Jr, Physics Essays 27, 591--7 (2014); DOI: https://doi.org/10.4006/0836-1398-27.4.591

\bibitem{pinheiro}
	M. J. Pinheiro, Physics Essays 20, (2007); DOI: https://doi.org/10.4006/1.3119404
	
\bibitem{covarMax}
	D. V. Redzic, Eur. J. Phys. 38, (2017); DOI: https://doi.org/10.1088/0143-0807/38/1/015602

\bibitem{lspeed2}
	D. Sadeh, Physical Review Letters 10, 271--3 (1963); DOI: https://doi.org/10.1103/PhysRevLett.10.271

\bibitem{sr4p}
	G. Stephenson and C.W. Kilmister, Special Relativity for Physicists Dover Edition (Dover, New York, NY, 1987)

\bibitem{waldron}
    R. A. Waldron, The wave and ballistic theories of light -- A Critical Review (Frederick Muller Limited, London, UK, 1997)

\bibitem{faster}
	W. D. Walker, Fundamental Theories of Physics (Springer, Dordrecht, DE 2002)

\bibitem{whit}
	E. T. Whittaker, A History of the Theories of Aether and Electricity (Harper, New York, NY, 1960)	
		
\end{thebibliography}
\end{document}